# Group-Based Trajectory Modeling of Citations in Scholarly Literature: Dynamic Qualities of "Transient" and "Sticky Knowledge Claims"

Susanne Baumgartner [a] and Loet Leydesdorff [b*]

**Abstract**
Group-based Trajectory Modeling (GBTM) is applied to the citation curves of articles in six journals and to all citable items in a single field of science (*Virology*, 24 journals), in order to distinguish among the developmental trajectories in subpopulations. Can highly-cited citation patterns be distinguished in an early phase as "fast-breaking" papers? Can "late bloomers" or "sleeping beauties" be identified? Most interesting, we find differences between "sticky knowledge claims" that continue to be cited more than ten years after publication, and "transient knowledge claims" that show a decay pattern after reaching a peak within a few years. Only papers following the trajectory of a "sticky knowledge claim" can be expected to have a sustained impact. These findings raise questions about indicators of "excellence" that use aggregated citation rates after two or three years (e.g., impact factors). Because aggregated citation curves can also be composites of the two patterns, $5^{th}$-order polynomials (with four bending points) are needed to capture citation curves precisely. For the journals under study, the most frequently cited groups were furthermore much smaller than ten percent. Although GBTM has proved a useful method for investigating differences among citation trajectories, the methodology does not enable us to define a percentage of highly-cited papers inductively across different fields and journals. Using multinomial logistic regression, we conclude that predictor variables such as journal names, number of authors, etc., do not affect the stickiness of knowledge claims in terms of citations, but only the levels of aggregated citations (that are field-specific).

---

[a] Amsterdam School of Communication Research (ASCoR), University of Amsterdam, Kloveniersburgwal 48, 1012 CX Amsterdam, The Netherlands; s.e.baumgartner@uva.nl .
[b] Amsterdam School of Communication Research (ASCoR), University of Amsterdam, Kloveniersburgwal 48, 1012 CX Amsterdam, The Netherlands; loet@leydesdorff.net ; * Corresponding author.

**Introduction**

Group-based Trajectory Modeling (GBTM; Nagin, 2005) provides a non-parametric statistics for distinguishing the developmental trajectories of subpopulations in sets. GBTM is based on using mixed models for the prediction of different trajectories in the data. This technique was first developed in fields such as criminology and clinical research in order to distinguish in an early stage, for example, youngsters who would be inclined to criminal behavior in a later stage of development (Nagin & Odgers, 2010a), or to predict the further development of symptoms and interventions in the clinic over time (Nagin & Odgers, 2010b). As against standard growth curve modeling, the group-based approach provides statistics for distinguishing among clusters of trajectories within a population (Andruff *et al.*, 2009; Nagin, 2005).

In this study, we explore GBTM by applying it to citation patterns in scientific literature. Citation patterns and citation windows can be expected to vary among fields of science (Price, 1972), among journals, and given co-variates such as document types or numbers of authors and pages (Garfield, 1979; Bornmann, Schier, *et al.*, 2012). In addition to the well-known impact factor, the *Science Citation Index* (SCI)[1] provides a number of journal indicators such as the immediacy factor, the cited half-life, etc., to trace such differences in development at the journal level over time. Furthermore, Thomson Reuters, the current owner of the Science Citation Index, provides *ScienceWatch* as an additional service: *ScienceWatch* lists fast-breaking papers at http://archive.sciencewatch.com/dr/fbp/. Papers which are immediately cited frequently can be considered as highly relevant and potentially ground-breaking (Ponomarev *et al.*, in press).

---

[1] We use "SCI" as a shorthand for the comparable databases for the social sciences (SSCI) and the arts & humanities (A&HCI). In this study, we use data from the SCI-Expanded version at the Web-of-Science (WoS).

Citation patterns tend to be heavily skewed, and are therefore far from normally distributed (Seglen, 1992). In recent years, the use of arithmetic averages (such as implied in the impact factors of journals) has been abandoned in favor of non-parametric statistics (Bornmann & Mutz, 2010; Leydesdorff *et al*., 2011; Waltman *et al*., 2012). Percentile distributions of citations can be used for calculating an "integrated impact indicator" (Leydesdorff & Bornmann, 2012); and increasingly consensus has emerged for considering the top-10% as an "excellence indicator" (Bornmann, de Moya-Anegón, and Leydesdorff, 2012; Waltman *et al*., 2012; cf. NSB, 2012).

However, the delineation of this top-10% is again dependent on the citation time-window, which may vary across journals, fields, and document types. The cut-off at the $90^{th}$ percentile is inspired by administrative standards (Bornmann & Mutz, 2010; NSB, 2012) rather than based on empirical evidence. The NSF, for example, distinguishes six classes (top-1%, 5%, 10%, 25%, 50%, and the remainder). However, it may also be that three (low, medium, high) or four classes (including a top group of most-highly cited outliers, or a never-cited group) are sufficient. In this paper we explore whether GBTM can provide an empirical solution to this problem. Is GBTM able to delineate empirically specific excellence and quality classes among citations in different journals and fields?

In a recent study, Ponomarev *et al*. (in press) focused on another indicator of ground-breaking research, "breakthrough papers". Breakthrough papers are identified by these authors at a high citation threshold (0.1% of the most frequently cited articles; cf. Ponomarev, in preparation) and thus as having a strong impact on the field. The citation trajectories of breakthrough papers as well as "excellent" papers are expected to show high levels of citations from the beginning; that is, they are immediately recognized as major contributions to the field. Citation rates of "excellent" papers are expected to remain high for a few years and then decline (Aversa, 1985; Price, 1976).

In contrast to these breakthrough papers, it has also been argued that the citations to "late-bloomers" (Merton, 1988) or "sleeping beauties" (Van Raan, 2004) emerge only gradually or after some years. These papers would follow a citation trajectory of no or few citations in the first years after publication, with a strong increase in citations after a few years. However, it is still unclear whether such "late-bloomers" are only exceptional cases that can be recollected from narratives, but are perhaps indicated idiosyncratically (Burrell, 2005). Is it possible to empirically identify a meaningful group of these "late bloomers"? GBTM may be a useful method to identify typical citation trajectories, as well as to detect more unique patterns such as "breakthrough papers" and "late bloomers."

In summary, GBTM may be of importance for the study of citation trajectories in the following ways:
1. Using GBTM, one is able to identify the typical shape of the development of citations over time. (After how many years do citations peak? When is the typical decline in citations? Is there a strong fluctuation or stability over time?).
2. GBTM makes it possible to identify subgroups of papers that follow specific citation trajectories (e.g., breakthrough papers or sleeping beauties).

3. GBTM can perhaps be used to identify excellent and breakthrough papers at an early stage.
4. GBTM allows us to compare citation trajectories across journals, disciplines, and scientific fields.
5. Using the statistical distinction between sub-populations, one can further ask whether external variables or co-variates (such as document and/or author characteristics) determine the likelihood that papers will follow a specific citation trajectory.

We explored GBTM by applying the technique to the citation curves of six journals in different research fields, as well as to citations in one entire research field (*Virology*) over a period of 16 years (1996-2011). We chose our samples to generate an overlap with three (of the eleven) papers studied as breakthrough papers by Ponomarev *et al*. (in press); these papers were published in 1996 in *Cell, Nature,* and *Science,* respectively. The other journals were chosen to explore potential differences across fields and/or conjectured similarities.

**Data**

We focus on six journals in different fields, using only articles published in 1996: the *Journal of the American Society for Information Science (JASIS),* the *Journal of the American Chemical Society (JACS), Cell, Gene, Science,* and *Nature*. Moreover, we chose one research field, *Virology*, including 24 journals. The choice of these six journals from among the 6,120 journals contained in the Journal Citation Reports for 1996, and the choice of *Virology* as a field, may seem somewhat arbitrary. As noted, the inclusion of *Cell, Science,* and *Nature* was motivated by the possibility to compare our findings with Ponomarev *et al.* (in press).

We chose additionally *JASIST, JACS,* and *Gene* as journals. *JASIST* is a leading journal in library and information science and provides us with familiar ground so that we can interpret aggregated patterns in terms of the underlying articles. The routine will first be explicated using *JASIST* as our lead example for the explanation. In 1996, *JASIST* was still named the *Journal of the American Society for Information Science* (JASIS) without the additional "*and Technology*" which was added only in 2001. The 1996-volume (vol. 47) contained 169 papers, of which we use only the 79 that are indicated as "research articles."

One of us compared *JASIST* and *JACS* (*Journal of the American Chemical Society*) in previous studies (Leydesdorff, 2001; Leydesdorff & Bensman, 2006). *JACS* provides us with a leading journal in one of the natural sciences, whereas *Cell* and *Gene* are typically biomedical. To broaden the focus of the study, we further investigated citation trajectories for the multidisciplinary journals *Nature* and *Science,* which also have a slight focus on the biomedical sciences.[2] Citation trajectories for multidisciplinary journals can be expected to differ from those for journals that focus on one specific research field.

Thirdly, we extend the study to a research field (i.e., *Virology*) operationalized as a WoS Subject Category. *Virology* was chosen for pragmatic reasons: we expected this field to be highly cited, sufficiently diverse, and relatively small. We retrieved the citable items in the 24 journals subsumed under this category in 1996: 110 review articles and 161 letters in addition to the 3,958 articles. One might expect reviews to be (significantly?) more frequently cited than research articles,[3] and letters in specialist journals less frequently than the average article (Leydesdorff, 2008, Figure 3 at p. 280).

---

[2] Using 13 categories, the journal list compiled for the US *Science and Engineering Indicators* series (NSB, 2012) by PatentBoards™, classifies *Science, Nature,* and *PNAS* as biomedical journals.

In the final step of the analysis, using mutinominal regression analysis, we investigate whether the type of publication as well as other co-variates (i.e., number of authors, number of references, page numbers and the journal name) can predict specific citation trajectories.

**Table 1**: Descriptive statistics of the sets under study.

|  | 1996 sets | Articles |
|---|---|---|
| *JASIST* | 169 | 79 |
| *JACS* | 2263 | 2142 |
| *Cell* | 466 | 346 |
| *Gene* | 760 | 747 |
| *Nature* | 3104 | 873 |
| *Science* | 2791 | 1064 |
| *Virology* (24 journals) | 4569 | 3958 articles; 110 reviews; 161 letters 4229 |

In sum, the selected journals allow us to compare similarities and differences in citation trajectories from different fields. At this stage, such an explorative approach seemed more informative than drawing a random sample from the journal domain. Table 1 provides an overview of the journals and numbers of articles in each journal. Of course, GBTM can be applied to all other journals and fields. This study is meant both as an explorative example of applying GBTM to citation curves and as a critical assessment of the usefulness of GBTM in citation analysis.

---

[3] 'In the *JCR* system any article containing more than 100 references is coded as a review. Articles in "review" sections of research or clinical journals are also coded as reviews, as are articles whose titles contain the word "review" or "overview".'; http://thomsonreuters.com/products_services/science/free/essays/impact_factor/ (retrieved on February 14, 2013).

**Methods**

Data was downloaded from the WoS interface in the second week of January 2013. Because citations to the 2012 volumes can be added until much later in 2013, we use citation data for the period 1996-2011, that is, 16 years. As noted, 1996 was chosen to facilitate comparison with Ponomarev *et al.*'s (in press) study of breakthrough articles, while at the same time keeping similar time lines across the sets in order to maximize the possibilities for comparisons among the cases.

GBTM has been developed as a subroutine in both SAS and Stata (Jones *et al.*, 2001). In this study, we use SAS 9.2 and the corresponding implementation PROC TRAJ, freely available at http://www.andrew.cmu.edu/user/bjones. Among the three available models, the zero-inflated Poisson (ZIP) model is most appropriate for citation count data, since one can expect more zeros than under the Poisson assumption (Lambert, 1992; cf. Hausman, Hall, & Grilliches, 1984).[4] The SAS syntax of the model is discussed in technical details in the Appendix.

Model selection is pursued in three steps. First, we stepwise increase the number of groups in the model specification. The Bayesian Information Criterion (BIC) is used as a test statistic for selecting the number of groups that best represent the heterogeneity among the trajectories. The BIC is known for penalizing over-fitting by introducing additional parameters. The selection of the model with the largest BIC is recommended,[5] but model selection should eventually be based also on domain knowledge and reasonable judgment (Nagin, 2005: 74-77). Furthermore, group sizes should be reasonably large (above 5%). In the first round, models with progressively more groups are tested until the model fit can no longer be improved.

---

[4] Rotolo, D., & Messeni Petruzzelli (in press) provide arguments why the negative binomial estimation is more appropriate for modeling citation data than assuming a Poisson distribution.

After identifying the number of groups, different shapes for the trajectories (linear, quadratic, cubic, etc.) can be tested in a second step. As the default, we assume that a citation curve can be expected to bend twice over a longer period of time, namely first rising to an apex of citations after two or three years (depending on the field of science) and then back again in the decline phase to an asymptotic approach to zero citations in the long run. We therefore defined all groups as following a cubic shape in the first step of the model fitting process. The shapes of the curves, however, can be adapted subsequently to alternatives that best fit the respective groups.

Once the ideal number of groups and shapes has been identified, in the third step, model adequacy can be tested using the average posterior probabilities (APP) of group membership. The posterior probabilities of group membership measure the likelihood for each scientific article to belong to its assigned group. Nagin (2005) recommends that the average posterior probabilities should exceed a minimum of .70 for each group. An average posterior probability of above .70 indicates that, on average, research articles are well assigned to their groups. In the graphs, 95% confidence intervals can be provided to show that confidence intervals of the identified groups do not overlap at specific time points.

---

[5] BIC is calculated as: $BIC = \log(L) - 0.5k \log(N)$. $L$ is the value of the model's maximum likelihood, $N$ is the sample size, and $k$ refers to the number of parameters in the model. In order to compare two models with different numbers of groups, the following estimate of the log Bayes Factor is used: $2\log_e(B_{10}) \approx 2(\Delta BIC)$ (Andruff, 2009; Jones et al., 2001; Nagin, 2005). In order to compute $\Delta BIC$, the BIC value of the simpler model is subtracted from the more complex model, and this value is thereafter multiplied by two. In accordance with recommendations of Jones et al. (2001), an estimated Log Bayes factor larger than five is considered as strong evidence for the more complex model.

**Results**

*a. Citation trajectories of JASIS articles published in 1996*

For the 79 articles published in *JASIS* in 1996, we tested models from one to seven groups: the BIC values of these models were -2318.69, -1773.22, -1638.90, -1619.49, -1608.57, -1605.32, and -1619.20, respectively. The six-group solution with all shapes defined as cubic therefore provided us with the best fit (BIC = -1605.32, log Bayes factor = 6.68). Since the six-group solution differs from the five-group solution only by distinguishing more subsets among the infrequently cited papers, we chose the five-group solution for presentation. The APPs for the five groups range from .92 to1.00, indicating that the research articles match excellently with their assigned groups.

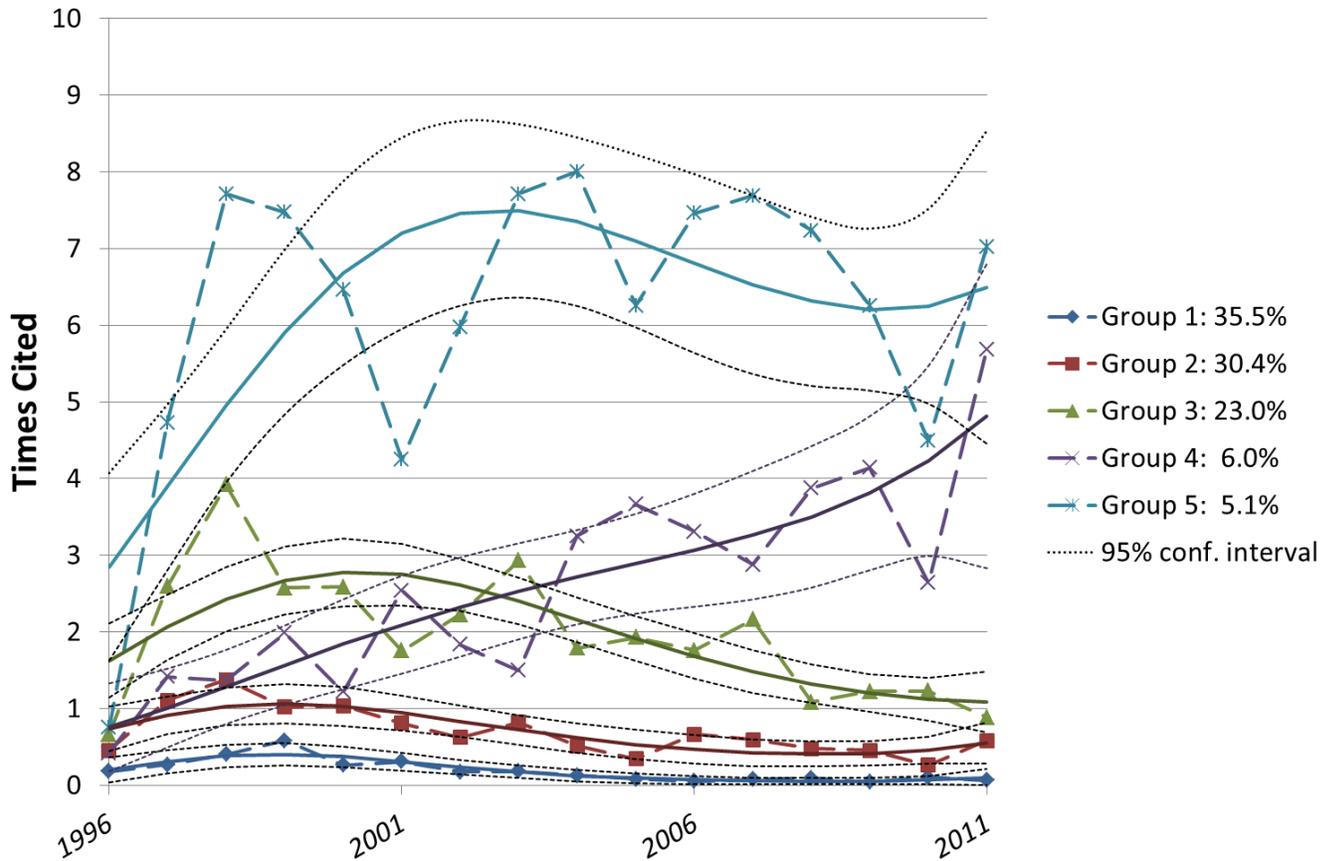

**Figure 1**: Five groups distinguished in terms of average citation rates among 1,517 citations to 79 articles from *JASIS* 1996, during the period 1996-2011. Smooth lines provide estimates; 95% confidence intervals are indicated with dotted lines.

Figure 1 shows the citation trajectories for these five groups during the 16 year time period. The five groups can be interpreted as follows:

1. A first (and largest) group consists of 35.5% of the papers ($n = 28$) which are almost never cited, that is, with an average citation rate below unity. Over time, these papers approach zero citations.
2. A second group of 30.4% of the papers ($n = 24$) is cited infrequently (approximately. once per year).

3. The third group comprising 23.0% of the papers ($n = 18.2$) is cited moderately ($< 4$). The citations in this group also seem to decline more slowly over time.
4. The fourth group (6.0%, $n = 4.7$) consists of papers that one could perhaps call "sleeping beauties." These papers are only infrequently cited in the first years after publication but their citation rates increase to an average of almost six citations in 2011.
5. Four papers (5.1%) are most frequently cited in this model; not surprisingly, these were also the most highly cited papers in the set of 79, with cumulative citation rates ranging from 78 to 132.

Note that the percentages provided in the legends are weighted in terms of the APPs of group membership and therefore the numbers per group do not have to add up to whole counts. Some cases cannot be attributed unambiguously to one group or another.

Interestingly, the five-group solution shows a "sleeping-beauty" pattern for Group 4. This pattern, however, was found only in the five-group model. In comparison, in the three- or four-group solution, a top group of *five* papers (6.3%) emerges, and the other papers previously attributed to the sleeping-beauty group are assigned to the moderately-cited group.

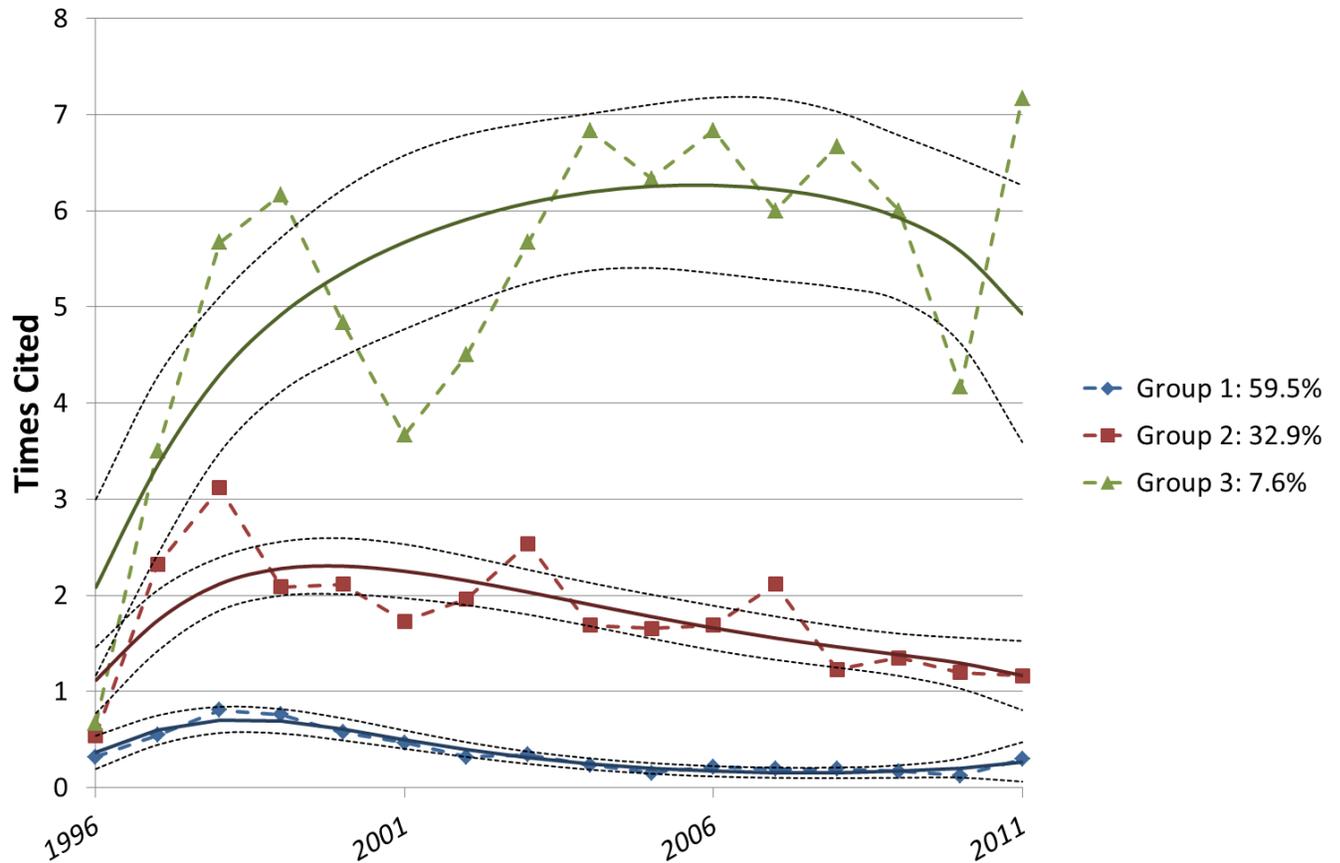

**Figure 2**. Three trajectories distinguished in terms of average citation rates among 1,517 citations to 79 articles from *JASIS* 1996, during the period 1996-2011.

Figure 2 provides the three-group solution in terms of most-highly-cited, medium-, and rarely cited papers. These three groups obviously need to be distinguished, in our opinion, since they are different already at the intercept and non-overlapping in terms of confidence intervals from the first year after publication.

In summary, the five-group solution plotted in Figure 1 is more precise, but Figure 2 provides an unambiguous separation into three groups. The highly-cited group in this (relatively small) set contained four or five out of 79 papers (5.1 or 6.3%, respectively). The five-group model also distinguished a sleeping-beauty trajectory. Given the small sample size, however, it seems premature to infer to the existence of this class of sleeping beauties. The percentage of most-frequently cited papers was considerably lower than 10% in all models.

*Journal of the American Chemical Society* (**JACS**)

In the following section, we present the models for *JACS*, *Cell*, *Gene*, *Nature,* and *Science*. Because the ZIP method of GBTM is sensitive to outliers, we found convergence failures in some of the cases. Citation distributions are extreme at both ends. The "zero-inflated" Poisson model (ZIP) accounts for the large numbers of zeros (non-citations) in the tails, but extreme values can also be expected at the high end. In the case of *JACS*, for example, the three most frequently cited papers (of 2142) were cited 2969, 2277, and 1594 times, respectively, with numbers four and five following at much lower levels, with only 783 and 670 aggregated citations. Under such conditions, GBTM fails to converge; we decided to consider these (three) outliers as another group to be excluded from GBTM.

The remaining 2,139 papers were included in the GBTM analysis. The BIC values in this case continued to increase when more groups were added until, with ten or more groups, the program failed to converge. Thus, the BIC value could not be used as a criterion for distinguishing the number of groups in this case. In such cases, Nagin (2005:74 ff.) recommends using domain knowledge to determine the number of groups.

Comparing the different models with one another, we found that when more groups are added, the model distinguishes mainly among the least cited papers. However, in the case of citation trajectories, one is most interested in the more frequently cited papers. Therefore, in the model selection process, we stepwise added more groups until no further meaningful distinction among the more frequently cited papers could be found.

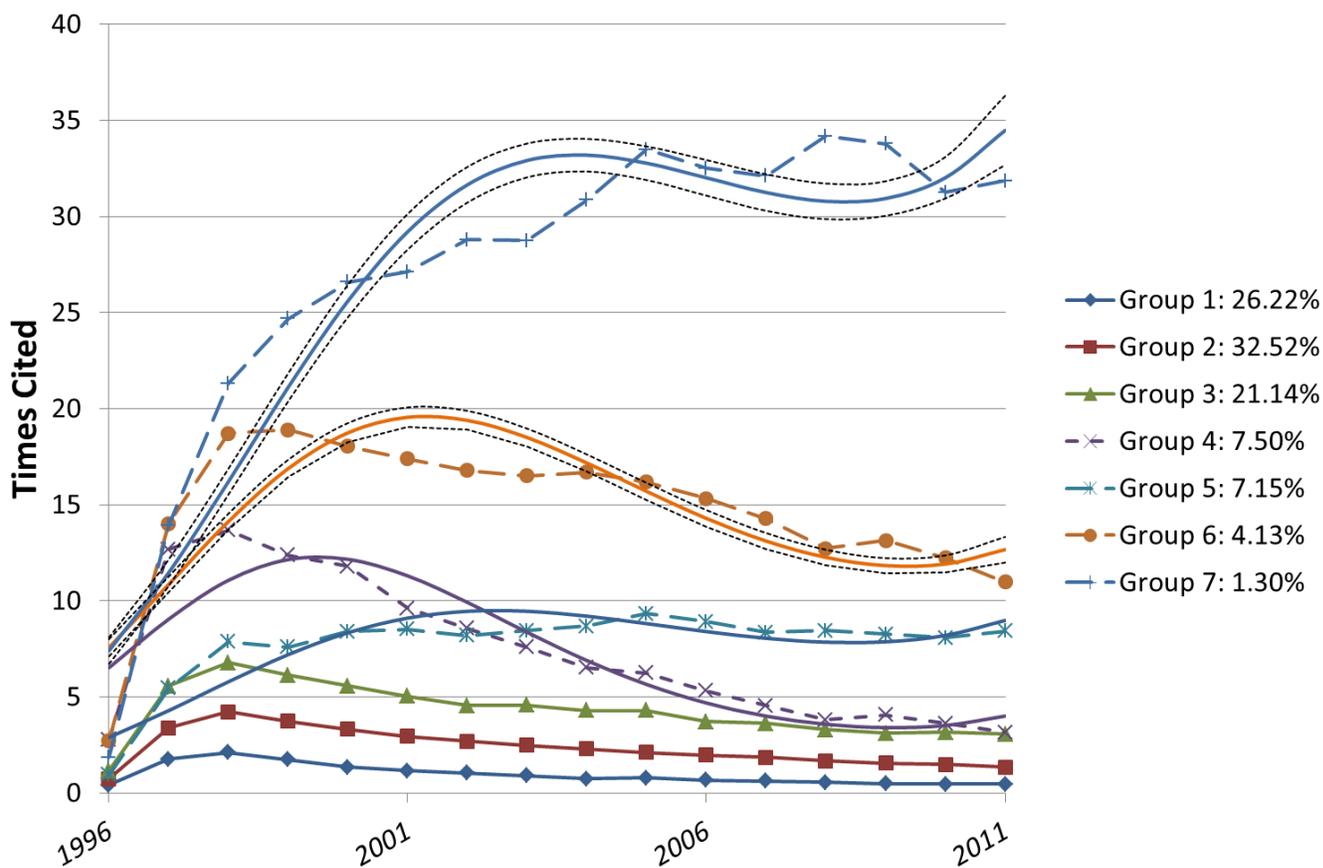

**Figure 3**: Seven trajectories of 2,139 articles published in *JACS* with publication year 1996.

After inspection of the possible models, we considered the seven-group solution as most informative (Figure 3). This model allows us to show that—from the top to the bottom—the first two groups (Group 7 with 1.3% of the papers and Group 6 containing 4.1%, respectively) are different in terms of continuing an initially similar (steep) increase in the number of citations. The curve of Group 6 reaches a peak level after a few years and declines thereafter. The topmost group, however, reaches a high level of citations and remains there. Although perhaps overcharging the terminology, one could say that these papers have become "citation classics" within this domain (and over a quite long period of 16 years).

The next two groups (Groups 4: 7.5%, and 5: 7.2%) differ similarly in terms of whether the citation curve bends back to asymptotically approaching zero or remains at a plateau through the entire period. Thus, we find this distinction both among the most-frequently cited papers and the moderately-cited papers. Perhaps, the absolute level of citedness can be considered as indicative of the intellectual fine-structure in this field in terms of respective specialties—e.g., organic, inorganic, and physical chemistry (Leydesdorff & Bensman, 2006; cf. Leydesdorff, 1991)—with different average citation rates (Garfield, 1979). Both in the highly-cited and the moderately-cited groups, however, a further distinction can be made between citation patterns that last for more than ten years and citation patterns that decay. We propose to name this difference in citation patterns as "sticky" versus "transient" knowledge claims.

For reasons of presentation, we did not add the confidence intervals for all groups in Figure 3. Inspection of the model depicted in Figure 3 made us aware that citation curves cannot properly be considered as following third-order polynomials, but that fifth-order polynomials would be more appropriate. Using fifth-order polynomials, the curves fit almost perfectly, with an explained variance of $R^2 > .95$ for all seven trajectories. This is shown in Figure 4. Using fifth-order polynomials in GBTM increases the BIC further to -72,323 as against -74,546 for third-order ones. The APPs range in this model between .93 and 1.00, indicating that the papers can be matched almost precisely into these seven groups.

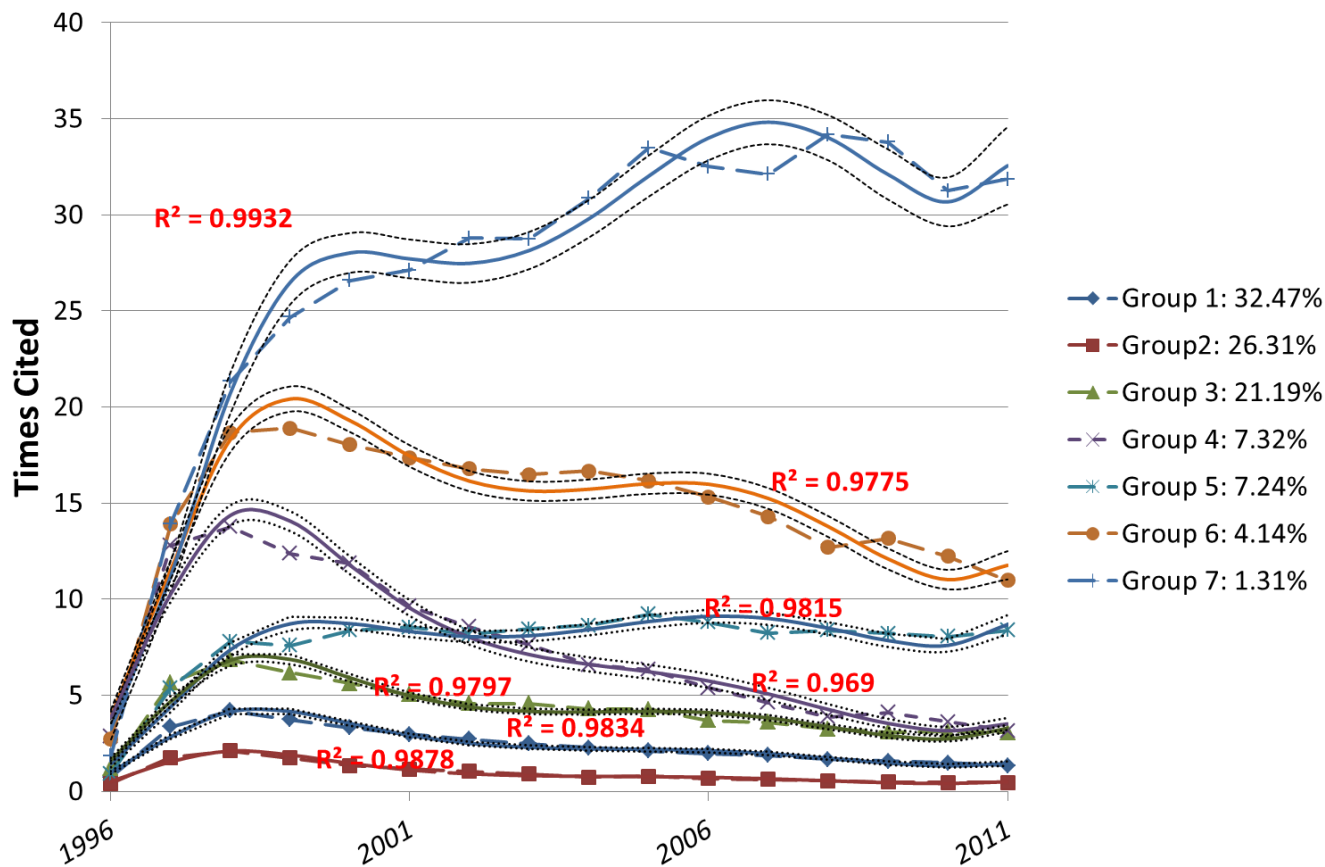

**Figure 4**: Seven trajectories of JACS using 5[th]-order polynomials.

Note that the size of the groups changed slightly when using fifth-order polynomials compared to the third-order polynomial groups. For example, the virtually non-cited group of papers is 26.31% in the case of using fifth-order polynomials and 26.22% using third-order polynomials. Further increases in the number of groups improved the BIC values in this case, but again this mainly affects the grouping of less-cited articles. In other words, it seems not possible to derive an optimal number of groups without making a qualitative judgment.

In summary, we found:

1. a number of subpopulations with different average levels of citations.

2. within the highly-cited and medium-cited groups, a further distinction between articles that have longer-term value ("sticky knowledge claims") and articles that typically function as references at the research front for only a few years ("transient knowledge claims"; Price, 1970). Further study of the less-cited groups also reveals this distinction at lower absolute levels.

3. a differentiated structure in the lower three subgroups that cover 80.0% of the articles: 26.3% remain almost uncited; 32.5% are cited incidentally; and 21.2% tend to remain at the level of four citations per year, and therefore can be expected also to contain "sticky" knowledge claims", but at a much lower level.

## *Cell* and *Gene*

Of the 346 articles published in *Cell* in 1996, the two most frequently cited papers belong to the group of outliers with 3,204 and 2,389 total citations, respectively; the third and fourth most highly cited papers had 1,848 and 1,830 total citations during this same period. For *Cell,* a three-group model fitted the data best (BIC = - 27,372.01). The BICs for the 4- and 5-group model were -27,392.45 and BIC = -27,412.90, respectively. Figure 5 shows these three groups;[6] the papers were perfectly assigned to the groups with all APPs being 1.00.

---

[6] The default starting values provided by SAS failed to find an adequate model. Therefore, starting values were specified in this case (see http://www.andrew.cmu.edu/user/bjones/example.htm for more information on this procedure).

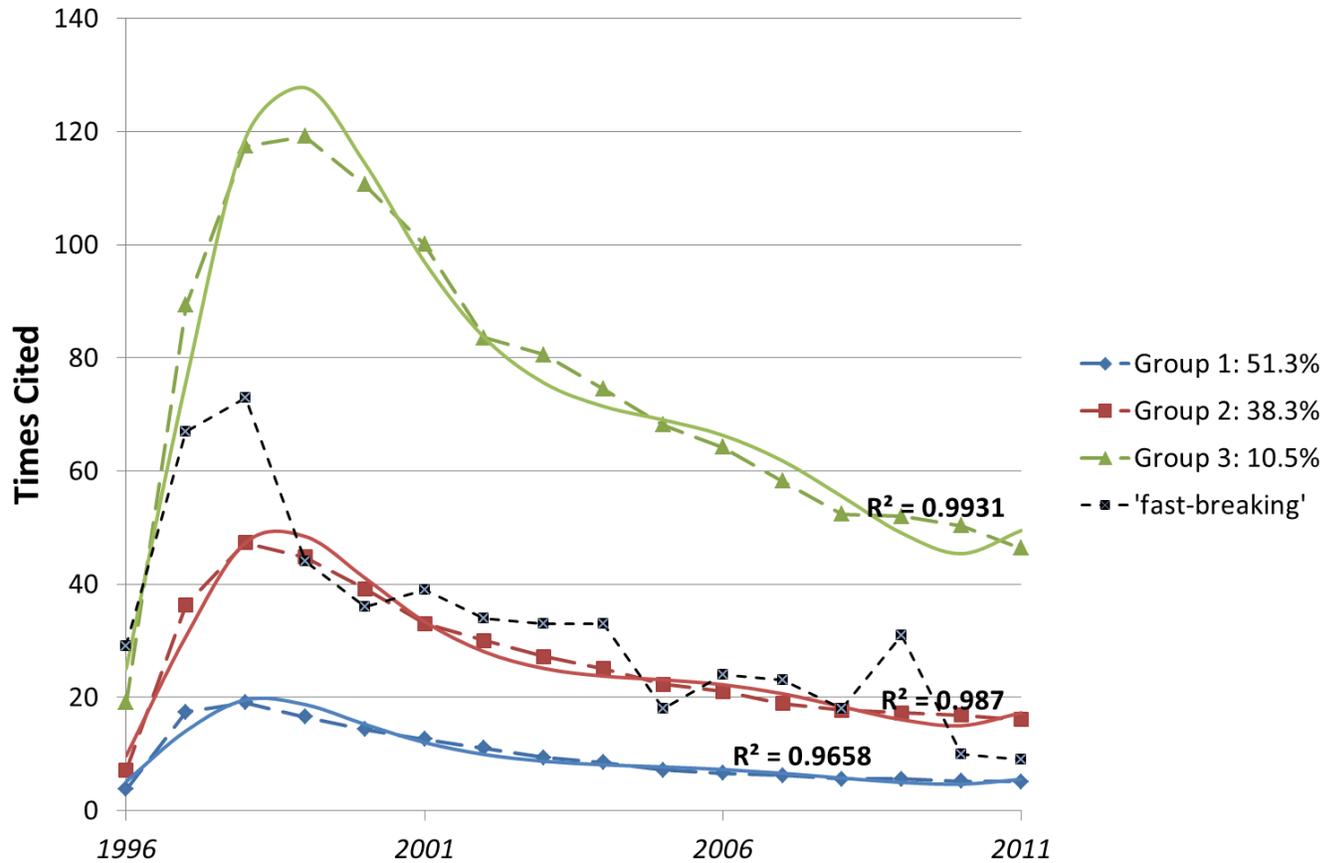

**Figure 5**: Three different trajectories on the basis of 344 articles published in *Cell* in 1996; 5[th]-order polynomials. The fast-breaking paper studied by Ponomarev *et al.* (in preparation) is added as a dashed line.

As noted, we chose the journal *Cell* because Ponomarev *et al.* (in press) included Hicke & Riezman (1996) as one of their set of "fast-breaking papers" in this year. However, this article is not part of the most highly cited group, but as shown in Figure 5 bends back in the second year after publication (1998) to a lower citation level, so that it becomes unambiguously (with a posterior probability of 1.00) part of the intermediate group.

Not surprisingly, the 5$^{th}$-order polynomials again fit with a very high level of precision ($R^2 > .95$) for all three groups. (Note that the fit is even $R^2 > 0.81$ for the single case of the fast-breaking paper.).

In contrast to *JACS*, we did not find a group of papers following a sticky citation pattern for *Cell*. Because *Cell* is a biomedical journal, it may be argued that the research front in the biomedical sciences moves faster than in other natural sciences. Perhaps citations decline faster in biomedical science than in the natural sciences, making a sticky citation pattern less likely.

To test this assumption we included an additional journal from the biomedical sciences in the analysis: *Gene*. *Gene* is in many respects comparable to *Cell* but is somewhat more specialized. Although citation rates were lower on average in this journal than in *Cell*, the most frequently cited group of papers (2.61%) convincingly shows a high degree of "stickiness" in their citation patterns (Figure 6). This indicates that also in the biomedical sciences, sticky citation patterns can occur.

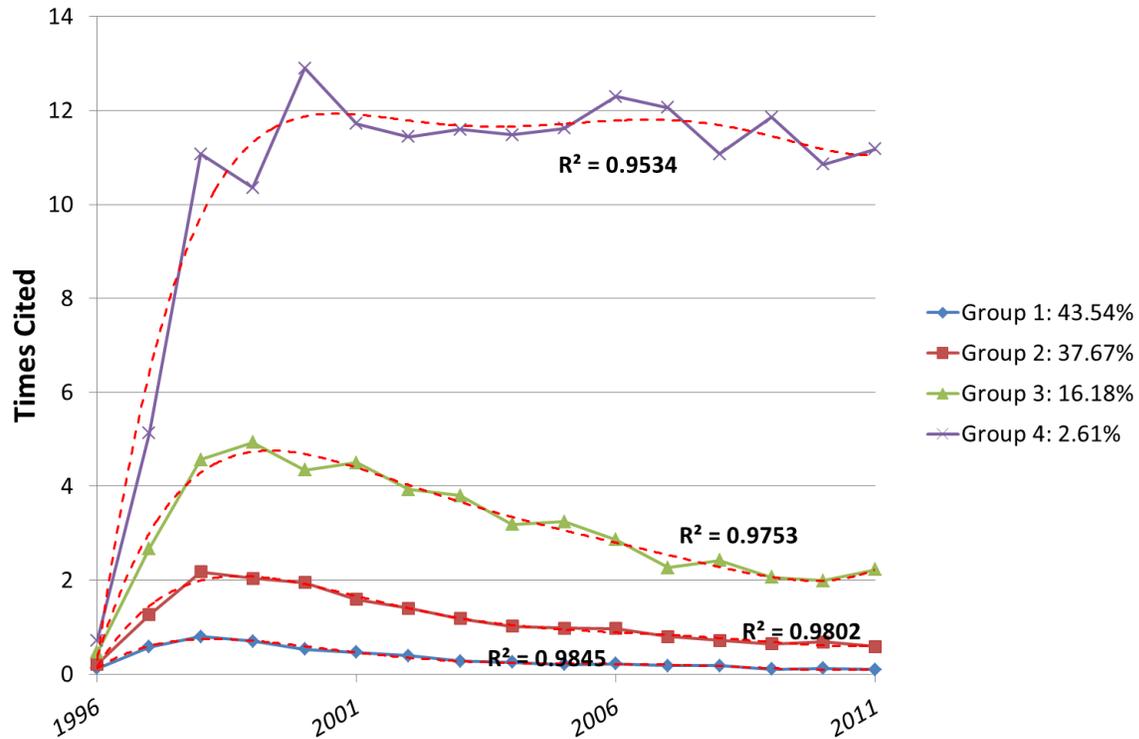

**Figure 6:** Citation trajectories of 746 cited articles, published in *Gene* during 1996. Only a single outlier had to be removed among the 747 articles published in *Gene* in 1996. GBTM with $5^{th}$-order polynomials failed to converge, but the $4^{th}$-order ones did. We added the trend lines with $5^{th}$-order polynomials to this figure using Excel.

*Nature*

Among the 873 articles published in *Nature* with publication year 1996, nine had to be removed as outliers in order to find a converging solution with GBTM. Four groups with $4^{th}$ order polynomials fitted the data best.[7] The BIC for this model is -60,734.15, all APPs are 1.00.

---

[7] Starting values had to be specified in order to find this optimal solution.

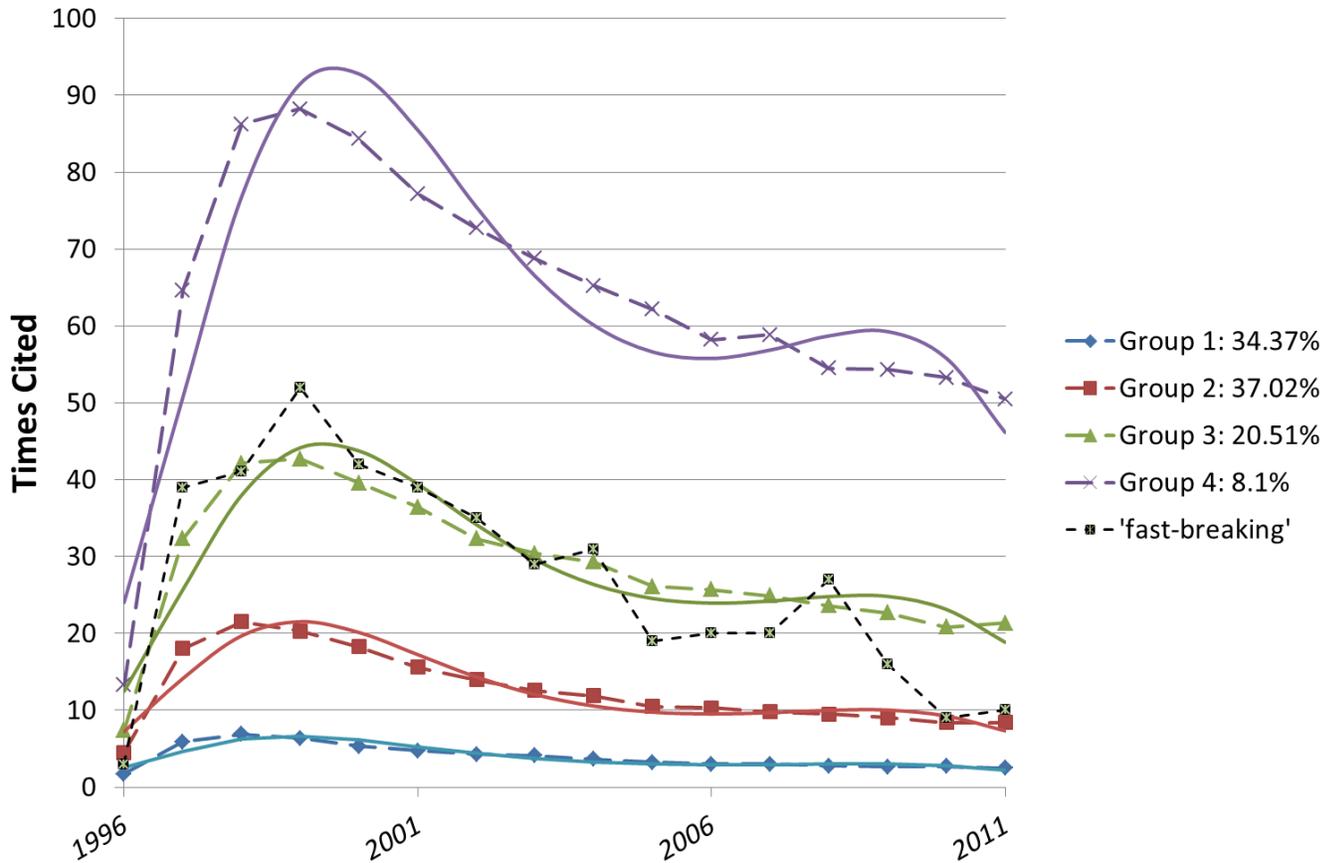

**Figure 7**: Four citation trajectories of 864 articles published in *Nature* during 1996; fourth-order polynomials.

In this case, the "fast-breaking paper" of Nussenzweig *et al.* (1996; see Ponomarev *et al.*, in press) is attributed unambiguously (APP = 1.00) to the medium-range group (Group 2); it peaks at 52 citations in 1999. Although this paper was fast-breaking immediately following its publication, GBTM shows that it does not belong to the most highly cited papers in *Nature* of this publication year. Furthermore, the decline in citations of this paper is rather striking, in contrast to papers that remain cited on a higher level throughout these years. In terms of cumulative total citations, it ranks only 174[th] among the set of 873 articles in the same journal (*Nature*) and with the same publication year (1996).

Comparison between Figures 5 and 7 (and also Figure 1) shows us the differences among fits when using different orders of polynomials. Polynomials with an order lower than five are not able to fit to the citation distributions because of the specific shape of the peak in the first few years. The fit with $4^{th}$-order polynomials in Figure 7 is improved when compared to the fit with $3^{rd}$-order polynomials in Figure 1, but the fit with $5^{th}$-order polynomials (in Figure 5) is precise. We discuss this issue in more detail in the discussion below..

*Science*

Among the 1,064 articles published in *Science* in 1996, 17 outliers had to be removed before we were able to find a converging solution. We used $5^{th}$-order polynomials to fit the remaining 1,047 articles. A five-group model emerged as the best fitting model. The groups are shown in Figure 8.

In this case, the "fast-breaking paper" included in the set of Ponomarev *et al.* (in press) belongs to the group of outliers among the reference set. (We included this group in Figure 8; as noted, it cannot be included into GBTM.) At its peak in 2002, this fast-breaking paper (Altman *et al.*, 1996) was the most frequently cited one in the set with 292 citations, but it exhibited the expected decline in citation rates in the years thereafter. Four other outliers obtained much higher citation rates in subsequent years (with approximately 600 citations per year for the highest ranking one). These four can be considered as "sticky knowledge claims" whereas the fast-breaking one followed a "transient" pattern.

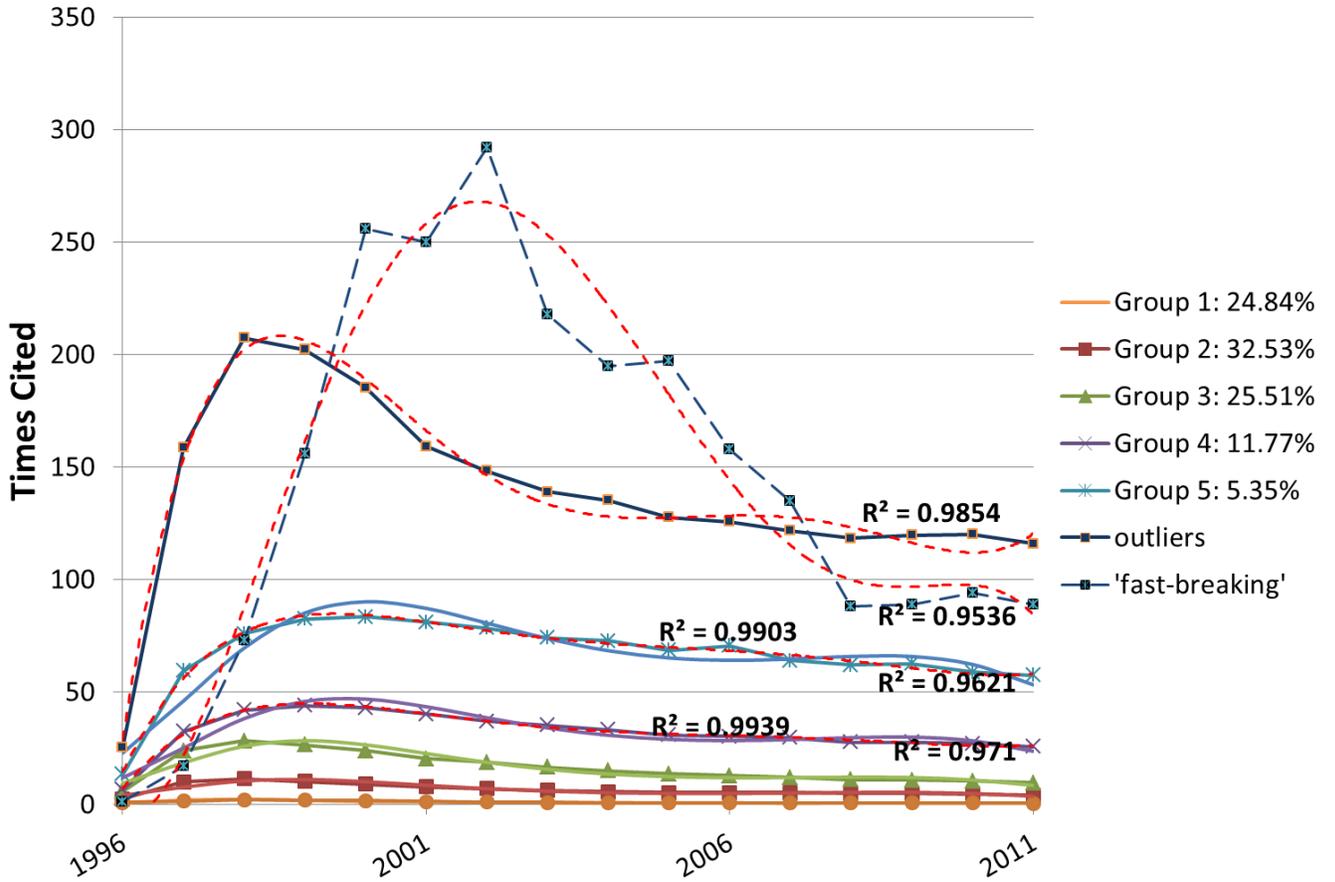

**Figure 8**: Citation trajectories of 1,064 articles published in *Science* with publication year 1996.

Figure 8 shows the excellent fits to the 5$^{th}$-order polynomials including a fit of $R^2 > .95$ for the single case of the "fast-breaking" paper. Interestingly, the overall decline in the tails of the distributions is less steep than in the case of *Nature*. Although *Nature* and *Science* are typically considered similar journals, these citation patterns may indicate that there are also differences. One reason for this difference in citation trajectories for the most highly cited papers may be that *Science* is less oriented towards the biomedical sciences than *Nature*.

*Virology*

Can GBTM also be applied to citation trajectories for a whole research field? We expanded the analysis to the research field *Virology* represented by 24 journals. In addition to articles, we included also reviews and letters as document types.

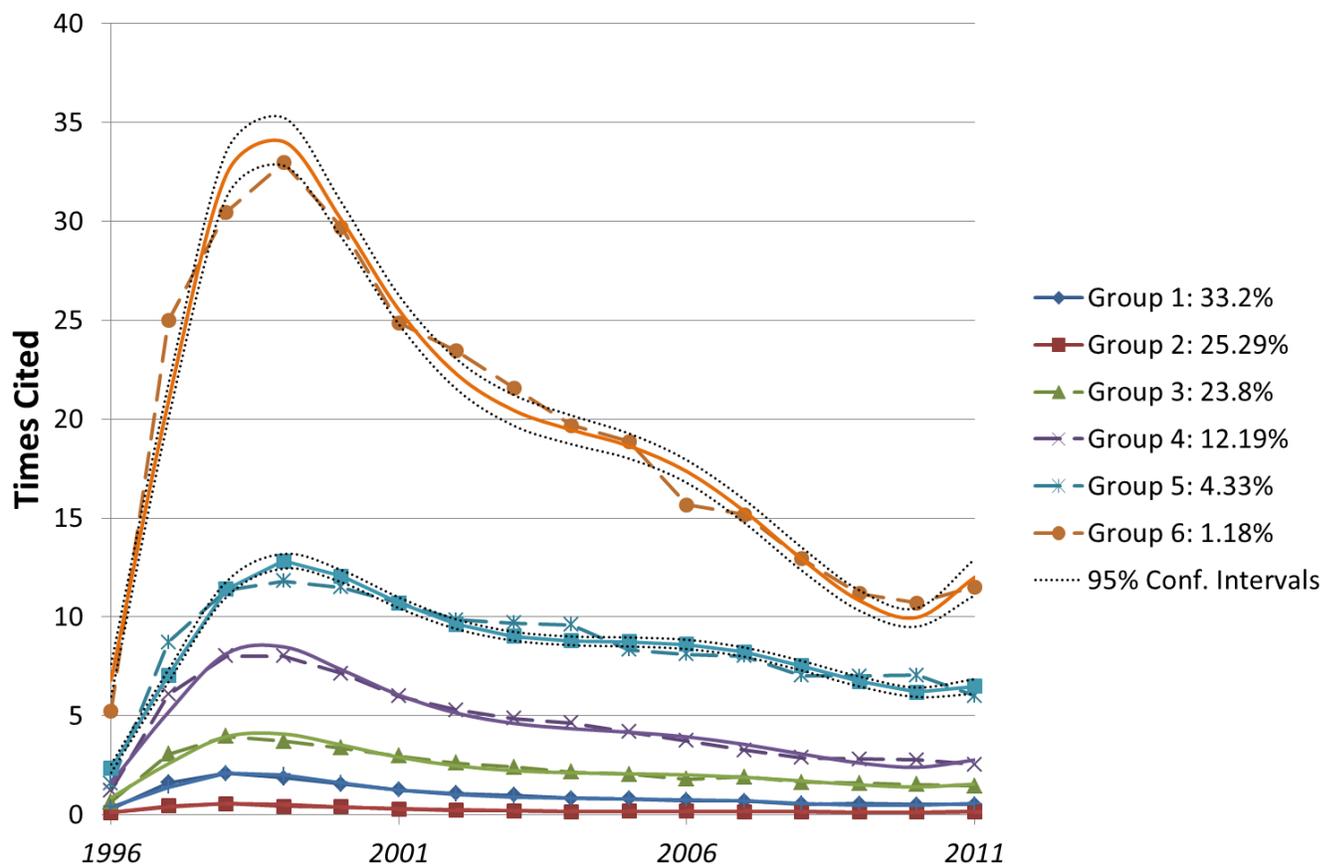

**Figure 9**: Six citation trajectories of 4,229 articles, reviews, and letters published in the WoS Subject Category of *Virology* with publication year 1996 (BIC = -107,949).

The choice of the number of groups remained a bit arbitrary also in this case, but the two top groups stabilize after six groups are distinguished. The six-group solution is depicted in Figure 9.

The two top groups consist of 1.18% ($n$ = 35.8) and 4.33% ($n$ = 183.1) of the papers, and—as before—these sets are much smaller than the "top 10%" of the Excellence Indicator. However, one can also reason that the top 10% would at least include all these excellent papers (Waltman *et al.*, 2012).

The six-group solution provided us also with an opportunity to compare these empirical results with the normative framework of the NSF (e.g., Bornmann & Mutz, 2010; NSB, 2012) that uses a scheme of top-1%, 5%, 10%, 25%, 50%, and bottom-50% in the rankings. The group sizes found using GBTM do not differ much from this distinction (as shown in Figure 9). Using the discrete classes defined by Bornmann & Mutz (2010), a strong overlap between our groups was found: $r$ = 0.87 ($p$<.05); $\rho$ = 1.00 ($p$<.01). If the top-5% of the NSF includes the top-1%, etc., using aggregation (NSB, 2012), our aggregated classes would be: 1.2%, 5.5%, 17.7%, 41.5%, 66.8%, 100.0%. The normative and empirical distributions of the classes are then virtually identical, with $r$ = 0.98 ($p$<.01); $\rho$ = 1.00 ($p$<.01).

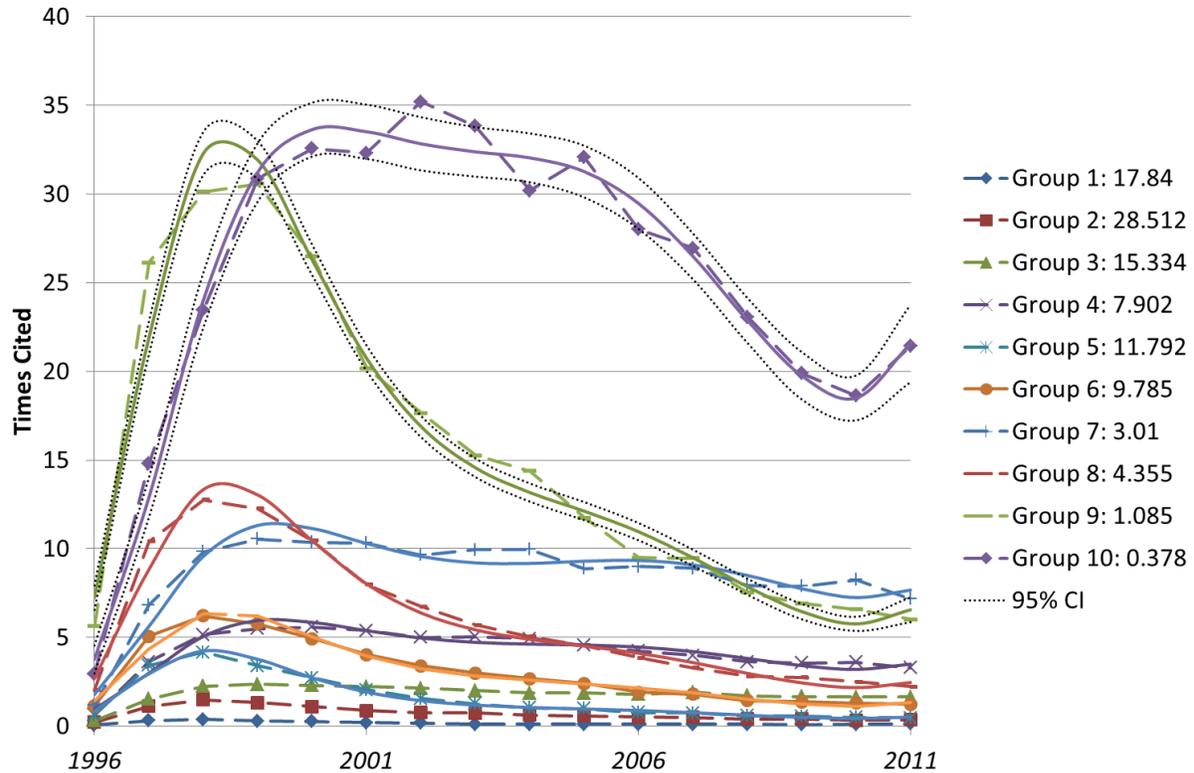

**Figure 10**: Ten citation trajectories among 4,229 articles, reviews, and letters published in the WoS Subject Category of *Virology* with publication year 1996 (BIC = -104,793).

If further groups are added to the model, we again find "sticky" and "transient knowledge claims." Figure 10 shows the model with ten groups. This fine-structure makes clear that both among the top-level papers 0.38% (*n* = 16 papers, Group 10) and among the fifth group in the middle range (3.01% or *n* = 334.2), articles with "sticky knowledge claims" can be distinguished from "transient knowledge claims." Similarly but at a lower level, Group 7 can be considered as exemplifying "sticky knowledge claims" whereas Group 8 shows "transient" ones.

When compared with Groups 8 and 9, the "transient" trajectories of Groups 7 and 10, respectively, peak earlier than the "sticky" knowledge claims. Whereas the "sticky" trajectories peaked only after four years, the "transient" groups peaked already after two years. This suggests that typical indicators of excellence which take only the first two years after publication into account fail to distinguish between these two potential pathways. In contrast to papers that follow a "transient" pattern, papers following "sticky" trajectories may have a more sustained influence on the research field and may therefore be better indicators of excellence independently of the absolute levels of the citations within each category.

**Co-variates**

The previous sections have shown that GBTM enables us to distinguish between different citation patterns over time. Most importantly, we saw that there are not only different levels of citations (in terms of total numbers of citations) but that one can distinguish also between "sticky" and "transient knowledge claims" within each level. In this section, we address the question of whether specific co-variates can predict which trajectories specific papers can be expected to follow. We show this by using the six groups distinguished for the *Virology* papers in the previous section (see Figure 9) as well as using the ten groups as presented in Figure 10.

As predictor variables we used document type (article vs. review vs. letter), number of authors, number of references, number of pages, and journal name (*Virology* is represented by 24 different journals). The variables at the interval scale (number of authors [NAU], number of references [NREF], and number of pages [NPG]; cf. Bornmann *et al*., 2012) were used as independent variables in a multinominal logistic regression with group membership as the dependent variable. In the first regression, group membership was based on the six-group model shown in Figure 9; the sixth group was defined as the reference group in the multinominal logistic regression[8]. The results are provided in Table 2.

The number of authors significantly predicted whether a document belongs to the 6th-group in comparison with almost all other groups ($p < .05$; only the difference between Group 5 and Group 6 was not significant at $p = .06$). The more authors an article had, the more likely the article belonged to the highest cited group. Furthermore, the number of references significantly differentiated the groups with lower levels of citations (Groups 1, 2, and 3) from the highest group. The more references an article had, the more likely it was to belong to the highest-cited group in comparison to the three lowest cited groups. However, the numbers of references did not significantly differentiate among the more frequently cited Groups 4, 5, and 6. The number of pages of a document did not significantly predict group membership.

---

[8] The equation for multinominal logistic regression is:
$$\log \frac{\Pr(group)}{\Pr(group6)} = a + b_1 X_1 + b_2 X_2 + b_3 X_3$$

**Table 2.** Parameter estimates of the multinominal logistic regression.

| Groups | | Estimate B | Std. Error | p-value | Odds ratio Exp(B) | 95% Conf. Interval for Exp(B) | |
|---|---|---|---|---|---|---|---|
| | | | | | | Lower Bound | Upper Bound |
| Group 1 | Intercept | 4.68 | 0.25 | 0.00 | | | |
| | NPG | 0.03 | 0.03 | 0.40 | 1.03 | 0.97 | 1.09 |
| | NREF | -0.03 | 0.01 | 0.00 | 0.98 | 0.97 | 0.99 |
| | NAU | -0.08 | 0.02 | 0.00 | 0.92 | 0.89 | 0.96 |
| Group 2 | Intercept | 6.07 | 0.27 | 0.00 | | | |
| | NPG | 0.00 | 0.03 | 0.95 | 1.00 | 0.94 | 1.06 |
| | NREF | -0.06 | 0.01 | 0.00 | 0.95 | 0.94 | 0.96 |
| | NAU | -0.19 | 0.02 | 0.00 | 0.83 | 0.79 | 0.86 |
| Group 3 | Intercept | 3.95 | 0.25 | 0.00 | | | |
| | NPG | -0.01 | 0.03 | 0.76 | 0.99 | 0.94 | 1.05 |
| | NREF | -0.01 | 0.01 | 0.03 | 0.99 | 0.98 | 1.00 |
| | NAU | -0.05 | 0.02 | 0.00 | 0.95 | 0.92 | 0.98 |
| Group 4 | Intercept | 2.94 | 0.24 | 0.00 | | | |
| | NPG | -0.02 | 0.03 | 0.53 | 0.98 | 0.93 | 1.04 |
| | NREF | 0.00 | 0.01 | 0.41 | 1.00 | 0.99 | 1.01 |
| | NAU | -0.03 | 0.02 | 0.04 | 0.97 | 0.94 | 1.00 |
| Group 5 | Intercept | 1.90 | 0.27 | 0.00 | | | |
| | NPG | -0.03 | 0.03 | 0.39 | 0.97 | 0.91 | 1.04 |
| | NREF | 0.00 | 0.01 | 0.82 | 1.00 | 0.99 | 1.01 |
| | NAU | -0.04 | 0.02 | 0.06 | 0.96 | 0.93 | 1.00 |

*Note.* The reference category is: Group 6. B is the regression coefficient. Exp(B) is the odds ratio of the predictor variable. Exp(B) indicates how the likelihood of an article belonging to the comparison group compared to the referent group (= Group 6) changes depending on the predictor variable. The model fit was adequate with Cox and Snell $R^2$ = .12 and Nagelkerke $R^2$ = .13. Model $\chi^2(15)$ = 552.61, $p$ < .001.

In a second multinominal logistic regression, the differentiation between ten groups was used as dependent variable, with Group 10 as reference group (see Figure 10). In this case, the co-variates no longer significantly predicted group membership, except that the number of references (NREF) among the lower-cited papers is significantly different from that of the most frequently cited group. In summary, these analyses show that the co-variates were not relevant for the distinction between transient and sticky knowledge claims as presented in Figure 10, but they are predictors for the levels and aggregates of citations (Bornmann *et al*., 2012) as is the case in the six-group model.

For the two categorical variables "journal name" and "document type", we conducted additional Chi-Square tests. These tests show that group membership to the six or ten trajectory groups significantly depended on the journal in which an article was published, $\chi^2 = 1,384.93$, $p < .001$ for the case of six groups, and $\chi^2 = 1715.49$, $p < .001$ for ten groups. For example, forty of the fifty papers in the top (sixth) group were published in the *Journal of Virology*. However, *Voprosy Virusologii* contributes with 78 papers and a total of 98 citations to the least-cited group and with two papers to the second lowest group (15 and 11 citations, respectively). Other non-English journals (*Bulletin de l'Institut Pasteur* and *Zentralblatt für Bakteriologie*) remain in the lower three groups, but journals like *Acta Virologica* and *Clinical and Diagnostic Virology* are also not cited above these levels. These journals thus seem to serve niche-markets.

The Chi-Square test for "document type" showed that significantly more reviews belonged to the most highly cited groups in comparison to letters and research articles, $\chi^2 = 231.20$, $p < .001$ for the six groups, and $\chi^2 = 242.54$, $p < .001$ for the ten groups. Letters were more likely to be attributed to the lower cited groups than the most-frequently cited ones. These results are in line with expectations.

Comparable to discriminant analysis, but in this case including not-normally distributed and categorical distributions, multinomial regression analysis also enables us to generate classification tables which cross-table predicted and observed group membership. Table 3 shows how well the respective co-variates predict group membership for the 6-group as well as for the 10-group model (Table 3).

| Predictor variable | % correctly predicted group membership (6 groups) | % correctly predicted group membership (10 groups) |
|---|---|---|
| Journal name | 41.0 | 32.4 |
| Document types | 35.0 | 29.8 |
| Nr of references | 37.1 | 32.5 |
| Nr of co-authors | 34.4 | 30.1 |
| Nr of pages | 35.5 | 30.6 |
| Times cited | 95.4 | 77.2 |

**Table 3**: Co-variates as independent predictors of group membership in the *Virology* set ($n = 4{,}229$; 24 journals; articles, reviews, and letters).

Not surprisingly, total "times cited" provides an almost perfect prediction of 95.4% in the case of six groups, but this is much less the case when ten groups are distinguished because—as shown above—in this case similar citation rates can indicate very different (sticky versus transient) citation patterns. The predictive value of journal name is rather strong with 41.0% for the six group solution and 32.4% for the 10-group model. The additional predictive value of the other co-variates is rather low. Adding these co-variates (document types, numbers of references and coauthors) to the journal names as predictors only improves the quality of the prediction from 41.0% to 44.2% in the case of six groups or from 32.4 to 34.5% in the case of ten groups.

In summary, this analysis teaches us that some of the co-variates can predict the number of citations in the aggregate, but do not allow us to distinguish between transient and sticky knowledge claims. Since the aggregated citation rates are often taken over the last few years for assessment purposes, the indicators tend to focus on transient knowledge claims.

**Discussion**

Before we turn to our conclusions, let us first critically discuss the usefulness of GBTM as a routine for identifying citation trajectories. The findings presented in this study show that GBTM can be applied successfully to citation trajectories. Although GBTM has previously only been used to measure the development of individual behavior over time, the present paper shows that it may also be applied very well to the citation trajectories of documents. The analysis advanced our knowledge of citation behavior by allowing us to describe different subgroups of articles that follow specific citation pathways over time. Instead of looking only at cumulative citations, GBTM enables us to delineate specific citation pathways.

However, the method also has limitations. One major problem remains the empirical specification of the number of relevant groups. In addition to using statistical parameters, it was also necessary to use more subjective assessments to select the best models. Another problem with GBTM is that the program sometimes fails to converge for complex models. Only by omitting outliers or by defining starting values could this problem be solved. Thus, the main use of GBTM remains its heuristic value: one can explore the data and become informed about the number of groups that should *at least* be specified. Therefore, GBTM can at present not yet be routinized instrumentally to classify large sets of data (e.g., for automatic application to the thousands of different journals in the database).

Although we had expected that a $3^{rd}$-order polynomial would fit the trajectories of citations, these polynomials showed a poor fit. Therefore, we exploratively tested higher-order polynomials. The analysis showed that $5^{th}$-order polynomials provided an excellent fit to the data. It seemed that these shapes adequately described the citation curves over the 16 year period. Surprisingly, the $5^{th}$-order polynomials fitted to curves on all levels, indicating that the shapes of the trajectories are in this respect similar for low and highly-cited papers.

Although higher-order polynomials will always lead to a better fit than lower-order ones, and $5^{th}$-order polynomials are extremely flexible so that the fit will easily be good, we were able to specify this effect in terms of the extremely rapid increase of citation curves in the first few years that leads to a sharp peak that cannot similarly be fitted otherwise. The fit of the $3^{rd}$-order polynomial misses this specific characteristic of the citation curve. The (analytical) difference between a plateau and a declining phase, and their possible mixtures in empirical cases makes it meaningful to account for more bending points in the curve than two.

One disadvantage of the excellent fits with $5^{th}$-order polynomials (four bending points) is that significance vanishes as a relevant criterion because the 95% confidence intervals become extremely narrow almost independently of the choice of the number of groups distinguished. Future research may also investigate whether the 5$^{th}$-order polynomials also fit the data well for shorter or longer time periods. For example, fewer bending points may be needed when using shorter time periods. Fitting the data to polynomials as is pursued when using GBTM, however, makes this method inappropriate for the prediction beyond the time-interval under study: the order of the polynomials is likely to determine whether the curves extrapolate to either zero or infinity in a relatively limited number of time steps (cf. Ponomarev, in preparation).

**Conclusions**

We explored the use of GBTM for distinguishing among the citation curves of differently cited documents. This study advanced our knowledge about citation curves in several respects. Most importantly, the analysis revealed two different citation pathways, which we named "sticky" and "transient knowledge claims". Papers that follow a sticky-knowledge citation trajectory continue to be cited throughout the years. These papers show a citation peak after three to four years after publication but the subsequent decline is less steep and these papers can still be highly cited after more than ten years.

Papers that follow a transient knowledge trajectory show a typical early peak in citations followed by a steep decline. After a couple of years, these papers are no longer frequently cited. These papers can be expected to fulfill a short-term function at the research front. The distinction between sticky and transient knowledge pathways was most apparent in the case of *JASIST*, *JACS,* and *Gene* (see Figures 1, 3, 4, and 6) and in the case of *Virology* (see Figure 10). Sticky and transient knowledge claims can be distinguished using GBTM only if sufficient groups are distinguished. Within the *Virology* set, for example, the two (analytically different) mechanisms remained entangled when six groups were distinguished, but became apparent when ten groups were declared.

Although one would expect "stickiness" to lead cumulatively to high citation rates in the long run, the focus on highly-citedness in the first two or three years induced by policy and management incentives has increasingly led to definitions of excellence in terms of "transient" knowledge claims. The sciences are different in terms of the extent to which a research front sets the agenda (Price, 1970), and some journals may function differently from others. It seems to us that this raises a number of follow-up questions for further research, such as whether dynamic features of citation curves should be introduced in performance and excellence measures. The failure of early prediction in terms of "fast-breaking papers" that was shown in the case of the papers in *Nature* and *Cell* shows that the quality of a paper in terms of citation impact cannot be concluded in the years shortly after its publication even in the case of a prevailing short-term research front.

In line with expectations, the identified citation groups differed not only in their citation patterns over time but also in a few other characteristics. In the case of *Virology*, multinominal logistic regressions as well as Chi-Square tests showed that the most highly cited group also differs from the lower cited groups in terms of the number of references cited within these papers, and the number of (co-)authors. Furthermore, citation patterns were also dependent on document types (letters vs. research articles vs. reviews) and the journals in which the documents were published (within the same field). However, these co-variates did not differentiate between papers following a "sticky" and "transient knowledge" trajectory. This indicates that these specific predictors can differentiate only between different levels of citations, but not different citation pathways. Future research is needed to investigate whether other indicators—e.g., institutional addresses—may enable us to predict these differences.

In a few cases, GBTM analysis also indicated a "sleeping-beauty" pattern. This was most apparent in the case of *JASIST* (see Figure 1, Group 4). To a lesser degree, this pattern was also found for *JACS* (see Figure 3, Group 5). This latter group showed a sticky pattern and at least some characteristics of "sleeping beauties". Why these sleeping beauties could be shown in the case of these two journals cannot be conclusively argued here due to the limited set of journals in this study.

Another finding of this study is that the citation curves over time seem to be more complex than expected. GBTM showed that $5^{th}$-order polynomials precisely matched the citation curves. The $5^{th}$-order polynomial provides this excellent fit to citation curves because in addition to the two bending points that we expected—convex at the apex and concave when bounding back in the decline phase—two more bending points can be expected in a potential plateau phase after reaching the top. In other words, the citation curve is not an exponential decay curve. The $5^{th}$-order polynomial also fits to the potentially steep increases of the citations in the first years.

Finally, this paper showed that the most frequently cited groups were in most cases much smaller than typical excellence indicators would predict. The most highly cited group consisted of 1% to 6% of the papers for the journals under study. This indicates that there might be fewer papers that should be defined as excellent than the previously discussed top-10%. GBTM is able to specify empirically the most frequently cited groups for each journal, but it is not possible with GBTM to define empirically an excellent group that holds across journals or fields. Journals and fields—and journals within fields—vary strongly in aggregated citation behavior, both in absolute terms—that is, as aggregates at each moment of time—and over time.

In sum, this paper introduced GBTM to citation trajectories. Despite some limitations of the method, GBTM provided us with new insights into the trajectories of citations over a longer time period. Most importantly, GBTM showed that citation curves are more complex and diverse than previously expected. By differentiating between sticky and transient citation patterns, the findings question typical "excellence" indicators that identify these papers in the first few years after publication.

## Acknowledgements

We thank Daniel Nagin, Lutz Bornmann, Ilya Ponomarev, and two anonymous referees for helpful comments on previous versions of this manuscript.

**Appendix**

The syntax in SAS to fit models with different numbers of groups and shapes can, for example, be formulated as follows:

```
PROC TRAJ DATA=off OUTSTAT=OS OUT=OF OUTEST=OE OUTPLOT=OP ITDETAIL;
ID id; VAR cit1-cit16; INDEP a1-a16;
MODEL ZIP; NGROUPS 5; ORDER 3 3 3 3 3; IORDER 2;
run;
%TRAJPLOT(OP,OS,'Citations vs. Time','Zero Inflated Poisson
Model','Citations','Time')
```

The first line specifies where SAS should write the output such as the file "OF" containing the posterior-probability attributions for all cases. The parameter "ITDETAIL" provides the value of the likelihood at each iteration. The variables involved are declared in the second line: the citation rates for 16 moments in time (cit1-cit16) given the respective time indicators (a1-a16) as independent variables.

"MODEL ZIP" in the third line specifies assuming the zero-inflated Poisson distribution; "NGROUPS = 5" asks for distinguishing five groups; and "ORDER 3 3 3 3 3" specifies the initial assumption of cubic equations for the fit of citation curves with two bending points. The parameter "IORDER" specifies the (linear or non-linear) function for the correction of additional zeros given the assumption of a Poisson distribution[9] The last line (%*TRAJPLOT*) asks SAS to plot the data using these legends for the axes. The various other parameters specify output files, such as "OP" containing the time-series data that can also be plotted using, for example, Excel. (The graphic interface of SAS is underdeveloped.) For a detailed tutorial on the model-fitting process using SAS, the reader is further referred to Andruff *et al.* (2009) and Jones *et al.* (2001).

---

[9] Because we ran into problems with memory requirements using 8GB, we decided to use the default for correcting the zero-inflation in the Poisson distribution instead of a non-linear correction (using "iorder 2"). Only in the case of *JASIST*, we present findings based on IORDER 2 but the difference between this model and the model based on the default value are marginal (behind the decimal point).